# Quantum teleportation with imperfect quantum dots

F. Basso Basset[1,*], F. Salusti[1], L. Schweickert[2], M. B. Rota[1], D. Tedeschi[1], S. F. Covre da Silva[3], E. Roccia[1], V. Zwiller[2], K. D. Jöns[2], A. Rastelli[3] and R. Trotta[1,*]

[1]*Department of Physics, Sapienza University of Rome, 00185 Rome, Italy*
[2]*Department of Applied Physics, Royal Institute of Technology, 10691 Stockholm, Sweden*
[3]*Institute of Semiconductor and Solid State Physics, Johannes Kepler University, 4040 Linz, Austria*
* francesco.bassobasset@uniroma1.it, rinaldo.trotta@uniroma1.it

**Efficient all-photonic quantum teleportation requires fast and deterministic sources of highly indistinguishable and entangled photons. Solid-state-based quantum emitters—notably semiconductor quantum dots—are a promising candidate for the role. However, despite the remarkable progress in nanofabrication, proof-of-concept demonstrations of quantum teleportation have highlighted that imperfections of the emitter still place a major roadblock in the way of applications. Here, rather than focusing on source optimization strategies, we deal with imperfections and study different teleportation protocols with the goal of identifying the one with maximal teleportation fidelity. Using a quantum dot with sub-par values of entanglement and photon indistinguishability, we show that the average teleportation fidelity can be raised from below the classical limit to 0.842(14). Our results, which are backed by a theoretical model that quantitatively explains the experimental findings, loosen the very stringent requirements set on the ideal entangled-photon source and highlight that imperfect quantum dots can still have a say in teleportation-based quantum communication architectures.**

## INTRODUCTION

Initially pursued as one of the most surprising consequences of the concept of entanglement in quantum information science[1,2], quantum teleportation has been increasingly investigated as a basic element for various applications in quantum technology over the course of the last couple of decades[3]. The photonic implementation of the protocol is essential to the exchange of qubits over long distances and in fully-fledged quantum networks[4,5].

Which technical platform has the better potential for bringing these applications to reality is still a question open for debate, even in the light of the impressive achievements reached with state-of-the-art solutions[6]. One of the main requirements concerns the generation of entangled photons[7]. In order to reduce losses and enable more complex functionalities, a deterministic photon source would be a fundamental asset, which is a major obstacle for probabilistic sources[8].

Semiconductor quantum dots (QDs) are a promising candidate for achieving on-demand operation. They can produce photon pairs in a nearly deterministic fashion[9,10] with extremely low multiphoton emission[11,12] and potential operation rates up to the GHz regime[13], even under electrical pumping[14,15]. As the light collection efficiency of the source[16,17] and the degree of entanglement[18] of QD-based photon sources is constantly improving, their performance is closing the gap with spontaneous parametric downconversion[19].

In recent years this series of achievements has led to seminal demonstrations of quantum teleportation using entangled QD photons to transfer qubits encoded in either an attenuated laser pulse[20–22] or a single photon generated on demand by the same QD[23]. While these results paved the way for a following generation of four-photon experiments[24,25], imperfections of the source still played a significant role forcing to either work at low values of protocol fidelity or with narrow temporal post-selection.

Dealing with these hurdles is especially relevant when considering the future roadmap for QD-based quantum communication, which inevitably points at remote QDs, posing additional demands on photon indistinguishability[26,27]. While these challenges keep motivating research on the QD fabrication, dealing with imperfect sources will certainly be mandatory in real-life applications. So far, little is known on the role of specific QD imperfections on the performance of quantum communication protocols.

In this work, we study in detail the impact that imperfections in the entangled photon source have on the success of the quantum teleportation operation. We investigate different schemes, comparing the simplest that selects only one out of the four Bell states to the one that implements a polarization-selective Bell state measurement (BSM). We find that these approaches not only differ in the protocol efficiency but, remarkably, also in the teleportation fidelity, by removing unwanted coincidences of distinguishable photons. Their performance can be further improved when moderate spectral filtering is applied. To show the relevance of our approach, we present our results using a *below-par QD*, i.e., one whose figures of merit (in terms of entanglement and indistinguishability) are below the average value found on the same sample. In this case, the average teleportation fidelity can be brought from below the classical limit—that is, failure of the protocol— to values as high as 0.84. The experimental results are supported by a detailed theoretical model that explains how the imperfections of the source can be mitigated by the choice of the protocol. Our work thus shows that the search for the *perfect QD* can be avoided and that the current quality of state-of-the-art QD entangled-photon sources is not so far from the more stringent requirements set by secure quantum communication applications[28].

## RESULTS

### Source and setup overview

Among the various available materials systems, our choice for the source fell on GaAs quantum dots grown by droplet etching epitaxy. The major advantage of this technology is the capacity to generate highly entangled states—up to 0.98 fidelity to a Bell state[18]—due to control over the in-plane symmetry of the nanostructure and to the fast recombination times with respect to the characteristic coherence time between the two bright exciton states[29,30]. The emission wavelength of emission can be chosen by fabrication design within the high-efficiency spectral region of femtosecond Ti:sapphire lasers and silicon-based avalanche photodiodes and fine-tuned with strain or external fields to match the operational requirements of Rb-based slow-light media[31–33] and quantum memories[34,35].

While, on the one hand, we make use of a state-of-the-art epitaxial technique, on the other hand, we do not resort to any strategy based on sample processing, external tuning, or performance-based selection to compensate for the unavoidable variability in the bottom-up fabrication process[36–38]. Therefore, we tackle the issue of source imperfections by considering a QD with suboptimal characteristics.

Figure 1a shows the emission spectrum of the QD studied in this paper, which consists of two polarization-entangled photons generated via the mechanism of the biexciton-exciton (XX-X) radiative cascade. Using a resonant two-photon excitation scheme, the photon pairs are produced with a preparation fidelity of 0.88, as estimated by intensity cross-correlation measurements[39]. We select this QD because the FWHM of the X and XX line is 118 and 68 µeV respectively, values that are more than a factor 10 larger than for the best QD in the sample. We attribute this to a particularly noisy QD environment that causes spectral diffusion at both short and long timescales[40]. We notice that the X line is broader than the XX line (a commonly observed feature for GaAs QDs[40,41]), but the difference in integrated intensity is less than 3%, suggesting that the role of XX relaxation processes alternative to the XX-X radiative cascade is negligible. The single-photon count rate is slightly above 1.5 MHz for an effective rate of laser pulses of 320 MHz.

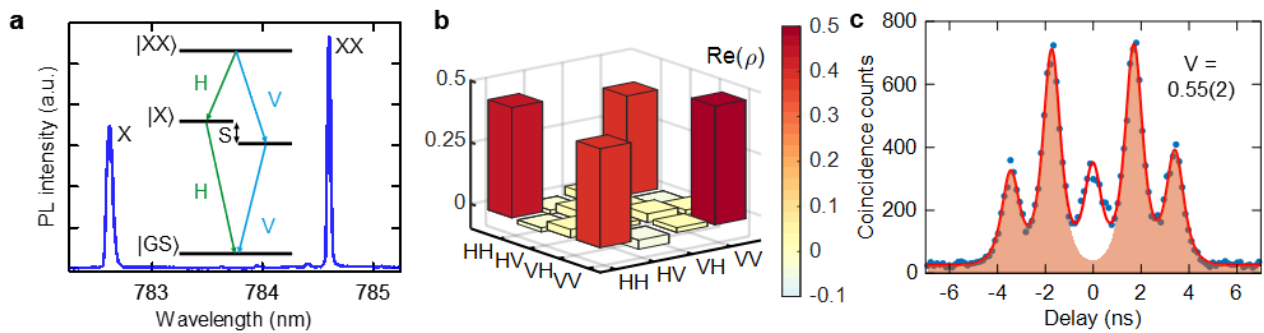

**Fig. 1 Main properties of the investigated QD. a** Photoluminescence spectrum of the investigated QD under resonant two-photon excitation of the XX state. In the central inset, the radiative cascade process is sketched with its energy diagram. **b** Real part of the experimental density matrix which describes the polarization state generated via the XX-X cascade in the rectilinear basis. **c** Intensity correlation histogram (blue dots) recorded in a HOM experiment on X photons with a relative emission delay of 1.8 ns. The data are fitted with the sum of five Gaussian functions convoluted to an exponential decay (red line), sharing the same FWHM. The fitted area of the side peaks is highlighted. The estimated interference visibility is also reported.

The features of the entangled photon source which are more relevant to the performance of the quantum teleportation operation are the degree of entanglement and photon indistinguishability[39]. In Fig. 1b we report the density matrix that describes the polarization state of the photons emitted through the XX-X radiative cascade. A liquid crystal retarder was set to maximize the fidelity to the state $|\phi^+\rangle = (|HH\rangle + |VV\rangle)/\sqrt{2}$, resulting in a matrix with no significant imaginary part (no matrix element above 0.03 in absolute value). The fidelity to the Bell state $|\phi^+\rangle$ is 0.89(1) and the concurrence of the matrix is 0.79(1), values well below the 0.98 and 0.97 that can be found in this materials system[18]. This is due to the presence of a finite fine structure splitting between the bright exciton levels, which is estimated to be 1.8(5) μeV. Next, we assess the photon indistinguishability of the X photons, which undergo the BSM in our teleportation scheme (see below), by performing two-photon interference in a Hong–Ou–Mandel (HOM) experiment. We select a single bright exciton component using a linear polarizer, so that the measurement is not influenced by the finite fine structure splitting. From the data reported and fitted in Fig. 1c, we estimate a visibility $V = 0.55(2)$. This figure is lower than the value of approximately 0.65 commonly reported for GaAs QDs with similar characteristics[26]. While the main reason for below-unity visibility is arguably the time correlation between photons emitted in a cascaded process[42,43], in this case a contribution from environmental field fluctuations is also probable, given the notable level of spectral diffusion in the time-integrated spectrum. Instead, we can safely assume no major reduction comes from the multi-photon emission of the source, because of the low values of the second-order autocorrelation function at zero-time delay $g_X^{(2)}(0) = 0.011(1)$ and $g_{XX}^{(2)}(0) = 0.020(1)$. The deviation from zero is mainly attributed to the incomplete suppression of the laser backscattering and to the afterpulsing effect of the detectors[44]. Having described the main features of our source, we now move to the experimental setup.

We build upon the quantum teleportation protocol employed in one of the seminal experimental demonstrations[2]. A qubit is initially encoded in the polarization state of a photon. This photon, together with one from a polarization-entangled pair, is projected with a measurement onto a Bell state, in such a way that its initial polarization state is transferred to the remaining photon of the entangled pair, after a unitary transformation which depends on the specific outcome of the BSM. Here, we specifically focus on the actual implementation of the BSM and improve on the basic setup employed in the recent realization of all-photonic quantum teleportation with deterministic QD-based sources[23].

Figure 2 summarizes the crucial parts of the setup. The protocol is performed using two pairs of entangled photons emitted by the same QD in two independent excitation events related to consecutive laser pulses, which we name early (E) and late (L) in their respective order. From the early photon pair, we only consider the photon from the X to ground state transition—namely the $X_E$ photon—which serves as the input state of the protocol. Its polarization is set using a linear polarizer (LP) followed by either a half-wave plate (HWP) or a quarter-wave plate (QWP). The late photon pair is spectrally split so that the $XX_L$ photon is sent to a polarizing beam splitter preceded by a QWP and a HWP, to perform quantum state tomography on the outcome of the teleportation protocol. The $X_L$ photon undergoes a BSM together with $X_E$. Its easiest implementation, shown in the upper right box of Fig. 2, relies on a non-polarizing beam splitter to observe two-photon interference. This allows for detecting only the antisymmetric Bell state, namely $|\psi^-\rangle = (|HV\rangle - |VH\rangle)/\sqrt{2}$, which caps the success rate of the protocol at 25%. The problem of how to realize a complete BSM has no easy solution[45,46], but the performance of the protocol can be improved with common linear optical elements[47,48].

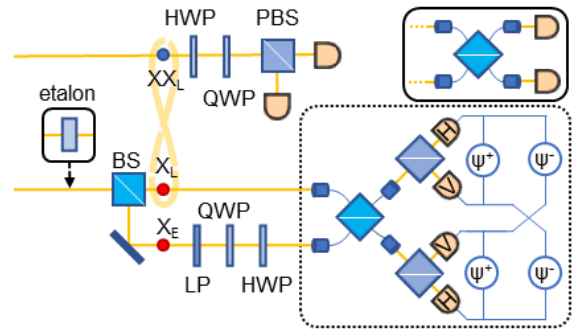

**Fig. 2 Quantum teleportation using a QD photon emitter.** Schematic illustration of the quantum teleportation setup. The BSM is implemented as shown in the dashed box, with a non-polarizing beam splitter (BS) followed by two polarizing beam splitters (PBS) at its output ports, allowing for the detection of two different Bell states. In the solid boxes we highlight the setup variations compared in this work. The detectors on the same side of the BS can be grouped to simulate the simpler procedure without polarization-selective elements. The X photons can be filtered with an etalon whose bandwidth is approximately twice the radiative linewidth of the QD emission.

The approach consists of adding two polarizing beam splitters at the output ports of the non-polarizing beam splitter, as illustrated in the main part of Fig. 2, and detecting photons from four channels instead of two. Considering pairs of detectors at the outputs of the same polarizing beam splitter, it is possible to measure the Bell state $|\psi^+\rangle = (|HV\rangle + |VH\rangle)/\sqrt{2}$ as well. At this point, one could naively think that these modifications only increase the success rate of the teleportation protocol to 50%, but this is not the whole picture when imperfections of the source are considered. In fact, the BSM setup with 50% discrimination probability does not only use photon indistinguishability to distinguish different Bell states, but also information about the polarization state. We will discuss and quantify this difference while presenting the results obtained with the different setups.

**Optimization by polarization-selective Bell state measurement**

First, we report the results of the quantum teleportation operation when the polarizing beam splitters in the BSM are disregarded, effectively mimicking the minimal setup with 25% efficiency sketched in the upper right box of Fig. 2.

Together with the coincident click of X photons exiting the non-polarizing beam splitter at different output ports within a time window of 0.6 ns, we record the polarization state of the $XX_L$ photon. For each input state, we measure the teleported state in the horizontal, diagonal, and circular polarization bases. In this way the polarization density matrix is derived from the Stokes parameters[49]. The total rate of threefold coincidences is 0.8 Hz, which is expected from the efficiency figures of merit of the setup and the source[39]. In Fig. 3a we report the output obtained when the input photon $X_E$ is sent through a linear polarizer oriented in the lab horizontal direction, selecting the H state. The first eigenvector is equal to its orthogonal state V, which is the expected outcome of the quantum teleportation operation when $|\psi^-\rangle$ is the measured Bell state, that is a bit-phase flip described by the Pauli matrix $\sigma_y$. The overall state is mixed and the fidelity to the V state is limited to 0.65(3). Figure 3b reports the teleportation fidelity for the diagonal (D) and right-circular (R) states as well, a set that allows evaluating the average gate fidelity for an arbitrary input[20], if the degree of polarization of the source is null. The average fidelity is equal to 0.644(17) and does not surpass the classical limit of 2/3, which corresponds to the threshold accuracy for the faithful cloning of a qubit[50].

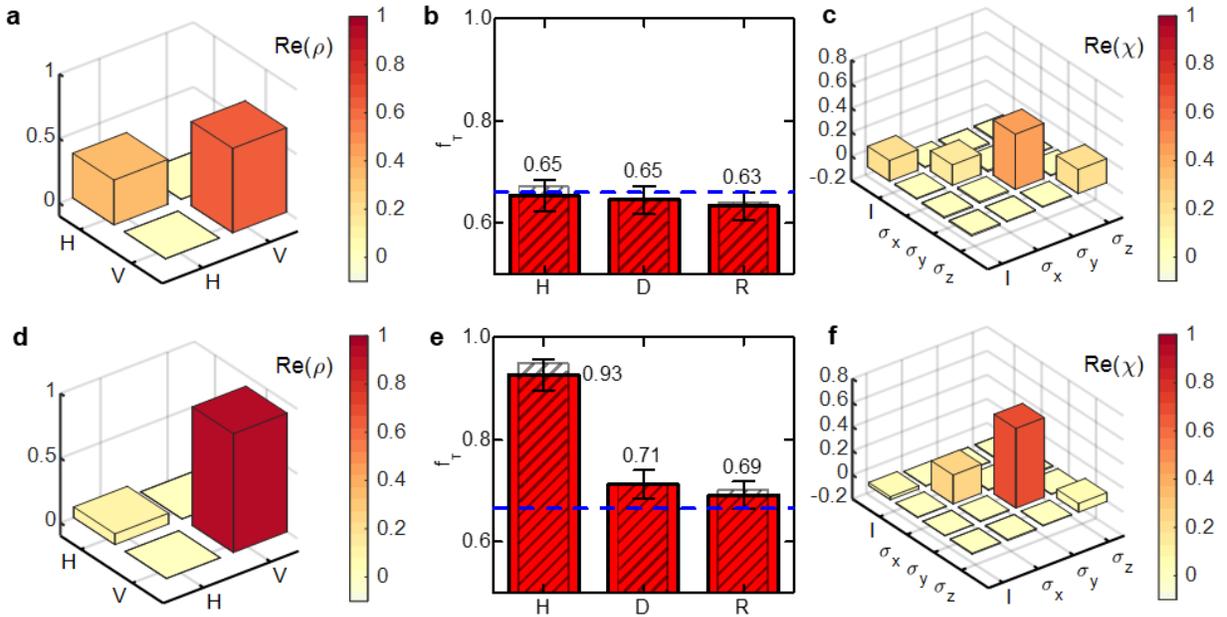

**Fig. 3** Quantum teleportation with different BSM implementations. **a, b, c** Measurement of the $|\psi^-\rangle$ state with a balanced non-polarizing beam splitter alone. In panel (**a**) density matrix of the teleported state corresponding to the input polarization state H. In (**b**) teleportation fidelities for the H, D and R input polarization states. The red filled bars represent the experimental values, the grey striped patterned bars are the values simulated based on the physical properties of the source. The blue dashed line represents the classical limit. In (**c**) quantum process tomography matrix. **d, e, f** Same as panels (**a**), (**b**), and (**c**) respectively, in the case in which the measurement of the $|\psi^-\rangle$ state includes polarizing beam splitters in the setup.

This conclusion is consistent with the source requirements set in previous works adopting the same BSM approach[23]. With an additional set of measurements we are also able to reconstruct the quantum process matrix[51] of the teleportation protocol. The result shown in Fig. 3c is a diagonal matrix whose elements correspond to the basic quantum operations $I$, $\sigma_x$, $\sigma_y$, $\sigma_z$, namely the Pauli matrices. These operations describe how the input state transfers to the output states in the ideal quantum teleportation for different outcomes of the BSM, $|\phi^+\rangle$, $|\psi^+\rangle$, $|\psi^-\rangle$, and $|\phi^-\rangle$ respectively. The process in Fig. 3c is essentially a mixed state of the Pauli matrix $\sigma_y$—the ideal operation for a $|\psi^-\rangle$ measurement, to which our matrix retains a 0.46 fidelity—together with the matrices related to the other BSM outcomes, present in similar proportions. From the quantum process matrix the average gate fidelity can be derived in a straightforward manner[52], yielding a value of 0.64, consistent with the estimate from Fig. 3b and still below the classical limit.

As mentioned above, the fidelity is expected to improve when we move to the polarization-selective BSM, i.e., when we consider the detectors corresponding to orthogonal linear polarization states as counting for a projection onto the Bell state $|\psi^-\rangle$. The difference is most striking for the input state H, as displayed in Fig. 3d. The purity of the output state is substantially increased, with a fidelity to V which is now as high as 0.93(3). The enhanced protocol also allows extending the operation to the detection of X photons in the Bell state $|\psi^+\rangle$. For each investigated input state, we observe that the discrepancy between the teleportation fidelities based on $|\psi^+\rangle$ and $|\psi^-\rangle$ detection lies within the statistical error (see the Supplementary Information for the complete comparison). In Fig. 3e we report the teleportation fidelities for different input states on the Poincaré sphere averaged over the different possible combinations of detection channels. The overall average teleportation fidelity is 0.776(16), more than six standard deviations above the classical limit. While we will discuss these results more quantitatively below, the quantum process matrix tomography for the measurement of the Bell state $|\psi^-\rangle$ in Fig. 3f already offers an intuitive picture of the main advantage of a polarization-selective BSM. Since coincidence events from distinguishable linearly co-polarized photons are discarded, teleportation events linked to the projection of X photons onto the Bell states $|\phi^+\rangle$ and $|\phi^-\rangle$ are removed. This can be noticed from the strong decrease in the matrix elements relative to the identity and to the Pauli matrix $\sigma_z$ with respect to the matrix of Fig. 3c. The increased purity of the process results in an average gate fidelity of 0.78 and 0.76 for the $|\psi^-\rangle$ and $|\psi^+\rangle$ case respectively (see the Supplementary Information for the matrix comparison).

Despite the strong improvement, we can object from Fig. 3e that the teleportation fidelity is still very close to the classical limit when sending D and R polarizations. Indeed, while classical correlation and sorting the polarization states of the X photons in the horizontal basis are sufficient to transfer the states H and V, entanglement and a selective BSM are essential to guarantee the protocol success regardless of the input state. In this case, the main limitation is clearly set by the limited photon indistinguishability of X photons coming from different radiative cascades.

**Optimization by spectral filtering**

Photon indistinguishability can be improved by means of spectral[53] or temporal[54] post-selection. Here, we study the effect of an air-spaced etalon with a bandwidth of 1.28 GHz, corresponding to 5.47 µeV. The introduction of this element obviously comes with a tradeoff in terms of source brightness. Given the large spectral diffusion of the emitter of this case study, this reduction amounts to a factor 10, which then corresponds to a factor 35 in the threefold coincidences.

However, the filter considered here is twice as large as the radiative linewidth of the X transition, so that the loss can be greatly reduced when investigating QD with linewidth closer to the Fourier limit. An intensity loss down to a factor 4 was indeed found for other emitters in the same sample.

Despite the modest selectivity of our filtering approach, its addition is greatly effective. The HOM visibility of X photons coming from consecutive radiative cascades is increased from 0.55(2) to 0.79(2) as displayed in Fig. 4a. This value is higher than the typical ones reported for similar GaAs QDs with the same excitation scheme and an X linewidth twice the Fourier limit[26]. The introduction of the etalon filter also has an impact on the wave packet of the emitted photons. From the time-dependent photoluminescence illustrated in Fig. 4b we can infer that the temporal width of the photon wavefunction is broadened because of the spectral filtering. We expect this temporal broadening to reduce the effect of the time-correlation in the cascaded emission process, as the effective temporal jittering with which the X state is populated becomes narrower relative to the temporal width of the photon wave packet. The detailed analysis of this effect is left to future studies.

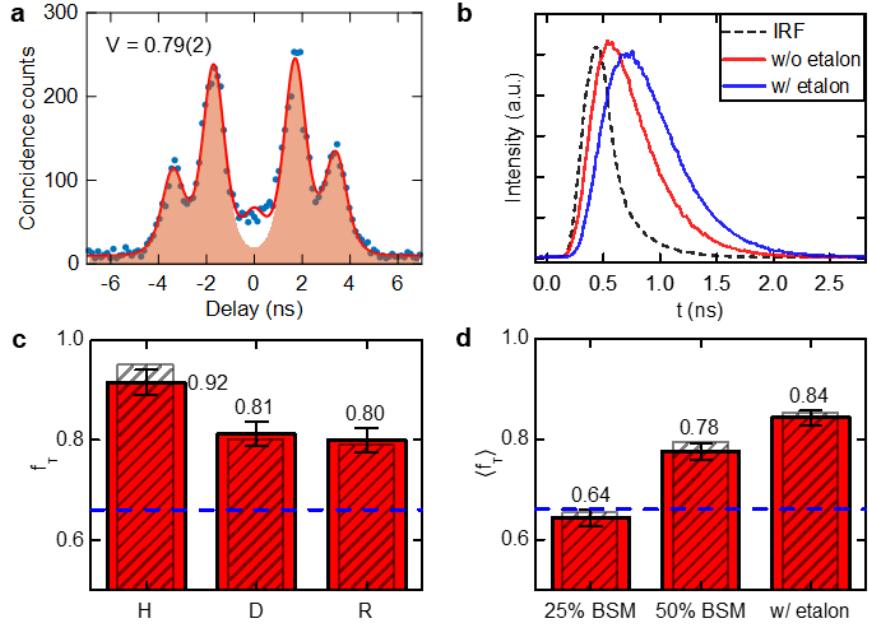

**Fig. 4 Quantum teleportation with spectral filtering.** Measurements performed on X photons transmitted through a 1.28 GHz bandwidth etalon. **a** Intensity correlation histogram (blue dots) recorded in a HOM experiment as in Fig. 1c. The data are fitted with the sum of five Gaussian functions convoluted to an exponential decay (red line), sharing the same FWHM. The fitted area of the side peaks is highlighted. The estimated interference visibility is also reported. **b** Time-resolved photoluminescence decay curves. Comparison between the case without (red curve) and with (blue curve) the presence of the etalon. The black dashed curve represents the instrument response function (IRF), which is convoluted with an exponential decay to fit the red curve and deduce the total lifetime of the X to ground state transition. **c** Teleportation fidelities for the H, D and R input polarization states. The red filled bars represent the experimental values, the grey striped patterned bars are the values simulated based on the physical properties of the source. The blue dashed line represents the classical limit. **d** Comparison of the average teleportation fidelity for the three approaches to quantum teleportation studied in this work.

Here, it is only important to notice that the change in the HOM visibility is not accompanied by a modification of either the photon emission statistics, with $g_X^{(2)}(0) = 0.009(1)$, or the degree of entanglement of the source, with a fidelity to the Bell state $|\phi^+\rangle$ of 0.90(1).

We repeated the teleportation experiment including the etalon filter at the entrance of the BSM apparatus. As expected, the higher HOM visibility of the X photons directly improves the success rate of the protocol. The results are reported in Fig. 4c. The teleportation fidelity relative to the input state H is essentially unchanged, whereas the values for the D and R input states are significantly increased with respect to the case of Fig. 3e and are now safely above the classical limit as well. This results in an average teleportation fidelity increased to 0.842(14), as evident from the comparison in Fig. 4d.

## DISCUSSION

In order to place the results reported so far into a quantitative framework, we introduce a formal description of the quantum teleportation operation for an imperfect entangled photon source. The protocol starts from the three-photon state $\rho_\Psi = |\Phi\rangle_{X_E}\langle\Phi|_{X_E} \otimes \rho_{X_L,XX_L}$, where $|\Phi\rangle_{X_E}$ is the input pure state to be transferred, and $\rho_{X_L,XX_L}$ is the density matrix representation of the polarization state of the photon pair emitted by the XX-X cascade. The latter matrix describes, in general, a mixed state, which can either be experimentally measured or simulated from the relevant figures of merit of the source, namely fine structure splitting $S$, multiphoton emission, and characteristic times $\tau_{SS}$ and $\tau_{HV}$ of decoherence between the bright X states[55]. The teleportation protocol acts on the three-photon state as a projection of the X photons onto a specific Bell state, e.g., $|\psi^-\rangle$, followed by a partial trace to return the teleported state:

$$\rho_{XX_L}^{\psi^\pm} = \sum_{i=\phi^+,\phi^-,\psi^+,\psi^-} p(i|BSM_{\psi^\pm}) Tr_{X_E,X_L}\left[\frac{\Pi_{X_E,X_L}^i \rho_\Psi \Pi_{X_E,X_L}^i}{N^i}\right]$$

(1)

where $\Pi_{X_E,X_L}^i$ is a projection operator onto the Bell state $i$, $N^i$ is a normalization factor, and $p(i|BSM_{\psi^\pm})$ is the probability that a Bell state detection event, here labeled

as $BSM_{\psi^\pm}$, is actually triggered by a photon pair in the Bell state $i$. The latter term accounts for inaccuracies in the BSM stemming from the non-ideal HOM visibility and fine structure splitting and, together with the expression for $\rho_\Psi$, describes the influence of QD imperfections on the teleported state. To make this point clear, consider the example in which two X photons are recorded at the separate sides of the non-polarizing beam splitter, and we tag the corresponding teleportation event as triggered by the detection of the Bell state $|\psi^-\rangle$. If the X photons are not fully indistinguishable, it is possible though that also photon pairs with a symmetric polarization state exit separately from the beam splitter. These accidental events are quantified in the non-zero values of $p(\psi^+|BSM_{\psi^-})$, $p(\phi^-|BSM_{\psi^-})$, $p(\phi^+|BSM_{\psi^-})$ that, via the sum in Eq. (1), enter in the final teleported state. It is also important to point out that $p(i|BSM_{\psi^\pm})$ is exactly the term that changes for the different setup implementations investigated in this work. Therefore, it is worth taking a look at how it depends on the HOM visibility and the fine structure splitting[39]. The dependence on the HOM visibility is summarized in Fig. 5a and 5b for the cases of the BSM with 25% and 50% efficiency respectively (see the Supplementary Information for the detailed formulas). Comparing the two graphs immediately highlights how the model treats different setup implementations of a $|\psi^-\rangle$ BSM. The inclusion of the polarizing beam splitters allows discarding with high precision coincidence events from linearly co-polarized photons, which means that the probability that a potential teleportation event is caused by X photons in a $|\phi^+\rangle$ or $|\phi^-\rangle$ state is negligible. This results in a dramatically increased accuracy of the BSM, especially when considering interference between photons with poor indistinguishability. It is worth noting that the HOM visibility is measured on a single bright exciton component and, therefore, does not include the effect of fine structure splitting on the photon indistinguishability. Even in the limit of ideal HOM visibility, a $|\psi^+\rangle$ state has a non-vanishing probability of being tagged as $|\psi^-\rangle$, since linearly cross-polarized photons are partially distinguishable if a frequency detuning due to fine structure splitting is present. This scenario is represented by the solid curve in Fig. 5a and 5b for the case in which the fine structure splitting is fixed to the value of 1.8(5) μeV of the investigated QD.

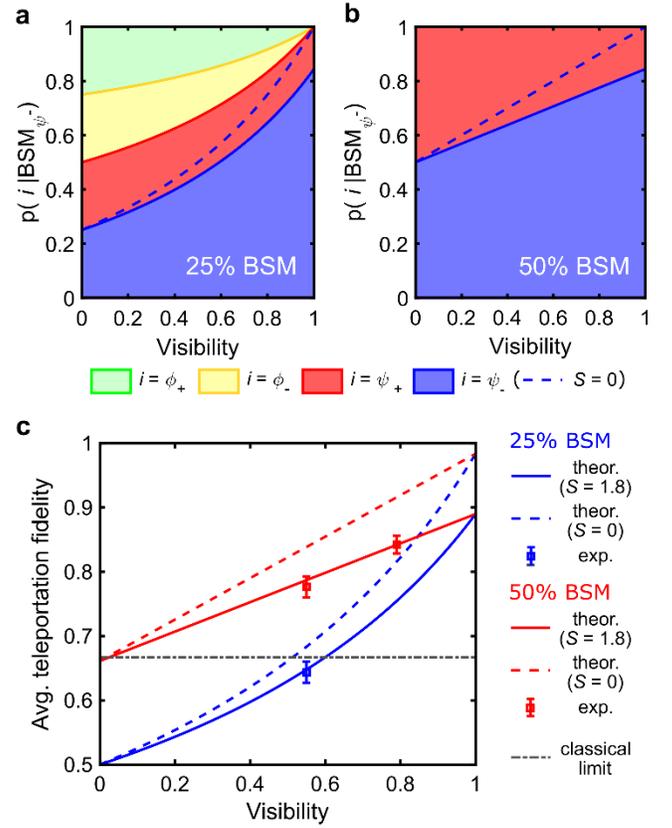

**Fig. 5** **Modeling with realistic source and setup. a** Probabilities that a BSM is caused by two photons in a given Bell state for the 25% efficiency setup, where the detectors at the same output of the non-polarizing beam splitter are grouped together. The probabilities are displayed in a stacked area plot as a function of the HOM visibility and are calculated both with all the physical properties of the investigated QD (solid curve, $S$ = 1.8 μeV) and assuming zero fine structure splitting (dashed curve). **b** Same as panel (**a**) but focusing on the BSM of the state $|\psi^-\rangle$ in the 50% efficiency setup. Identical results apply to the state $|\psi^+\rangle$ by simply swapping the labels between the two states. **c** Simulated values of the average teleportation fidelity as a function of the HOM visibility for both the implementations of the BSM. Calculated both with all the physical properties of the investigated QD (solid curve, $S$ = 1.8 μeV) and assuming zero fine structure splitting (dashed curve). The graph also presents the experimental average fidelities (square dots) from Fig. 4d for comparison.

We can evaluate these conditional probabilities for the QD of interest from a few measurable physical properties of the source, namely HOM visibility, fine structure splitting, X radiative lifetime and pure dephasing rate $T_2^*$. For the latter parameter, we adopt a literature value from a comparable GaAs QD[40], as its precise estimation only marginally affects the predictions of the model. Knowing the experimentally measured density matrix $\rho_{X_L,XX_L}$, we are able to simulate the teleported state for each experimental condition and input polarization analyzed in our work. The teleportation fidelities relative to the simulations are indicated by grey striped bars in Fig. 3b, 3e, 4c, and 4d. The excellent agreement with the experimental data fully supports our previous insights on the role of the polarization-selective BSM and the spectral filtering on the success rate of the quantum teleportation protocol based on photons from a quantum emitter.

The model also qualifies as a useful tool in guiding the design of a source useable in a realistic quantum communication protocol. By including a formula for the entangled pair density matrix as well, it is possible to predict the protocol performance from a few physical properties of the source alone. We add a minor variation with respect to the matrix introduced by Hudson et al.[55] (see the Supplementary Information for the complete expression), by also accounting for a unitary transformation applied on both photons to maximize the fidelity to the Bell state $|\phi^+\rangle$, which is done in our setup with a liquid crystal retarder.

In this way, we obtain an analytical expression for the quantum process matrix $\chi$ which provides an exhaustive description of the teleportation procedure[51]. $\chi$ is a diagonal matrix, whose expression depends on the specific outcome of the BSM, in this case either $|\psi^-\rangle$ or $|\psi^+\rangle$

$$\chi_{00}^{\psi^\pm} = \chi_{33}^{\psi^\pm} = \frac{1 - c' k g_{H,V}^{'(1)}}{4}$$

$$\chi_{11}^{\psi^\pm} = \chi_{22}^{\psi^\mp} = \frac{1}{4}\left(1 + c' k g_{H,V}^{'(1)} \pm c \frac{2 k g_{H,V}^{(1)}}{\sqrt{1 + \left(\frac{S\tau_0}{\hbar} g_{H,V}^{(1)}\right)^2 \left(1 + \left(\frac{S\tau_0}{\hbar} g_{deph}^{(1)}\right)^2\right)}}\right)$$

where $k$ is the fraction of detected photons pairs in which both photons come from the XX-X cascade and not from multiphoton or background emission, $S$ is the fine structure splitting, $g_{H,V}^{'(1)} = 1/(1 + \tau_X/\tau_{SS})$, $g_{H,V}^{(1)} = 1/(1 + \tau_X/\tau_{SS} + \tau_X/\tau_{HV})$, and $g_{deph}^{(1)} = 1/(1 + 2\tau_X/T_2^*)$. The coefficients $c$ and $c'$ depend on the specific implementation of the BSM, either with (BS) or without (PBS) polarization-selective elements:

$c'_{BS} = \frac{V}{2-V}$, $c'_{PBS} = 1$ and $c_{BS} = \frac{V}{2-V}$, $c_{PBS} = V$.

The fine structure splitting causes quantum teleportation to act as a mixture between $\sigma_x$ and $\sigma_y$ single-qubit gates, assuming that the symmetry axes of the QD are aligned with the horizontal polarization basis of the lab. Its impact is reduced by using a polarization-selective BSM except in the case of otherwise perfect photon indistinguishability. The effect of the BSM implementation is more drastic the lower the HOM visibility of the emitter. If we only rely on photon indistinguishability to sort out the Bell state, an imperfect overlap of the wavefunctions causes the ideal process matrix to mix with an identity matrix, the same effect as spin scattering mechanisms that erase any correlation in the polarization of the entangled photon pair. If we also select Bell states with the help of polarizing elements, the effect is limited to mixing the $\sigma_x$ and $\sigma_y$ elements of the process matrix, as for fine structure splitting and cross-dephasing mechanisms, and not to the elements related to the discarded $|\phi^+\rangle$ and $|\phi^-\rangle$ states.

From the quantum process tomography an expression for the average fidelity of teleportation can be easily obtained as well. In Fig. 5c, we show how the fidelity of teleportation changes with respect to the HOM visibility for both the QD investigated here and a similar one with its fine structure splitting brought to zero. The curves are compared once again with the experimental data, confirming the excellent agreement.

In summary, we have investigated in detail how imperfections in a QD source affect the performance of quantum teleportation using photons and their polarization state, in a way that could be easily extended to any generic polarization-entangled photon source relying on a radiative cascade emission process. The fine structure splitting and, even more so, two-photon interference visibility may not be optimal in as-grown QDs and influence the probability of success of the protocol. However, their impact on the fidelity of the operation strongly depends on the implementation of the BSM in the setup. A direct comparison by means of quantum process tomography highlights that the inclusion of polarizing beam splitters, in addition to increasing the efficiency of the protocol, allows sorting out Bell states with higher accuracy and lowers the source requirements for quantum teleportation. This approach has also been studied in combination with spectral filtering. The choice of a relatively large bandwidth—double the radiative linewidth—leads to a significant increase of the HOM visibility, in the test case from 0.55(2) to 0.79(2), while not affecting in a relevant way other important features such as photon statistics and degree of entanglement. Altogether, the average teleportation fidelity is raised from below the classical limit to 0.842(14). Finally, we have presented and discussed a quantitative model able to predict the performance of the protocol from the physical properties of the source and the details of the setup, with the goal of providing a useful tool for experiment design. Recent improvements in terms of sample fabrication and processing, such as integrated micromachined piezoactuators[36,56] and broadband photonic cavities[16,17], hold great promise of tackling the limitations of existing quantum emitters at their roots. However, the emerging paths of development in the field—above all quantum communication over long distances[22,57] and between remote emitters[26]—come with additional technical challenges, so that managing source imperfections will remain a crucial task for reaching the next milestones. Therefore, we envisage that the strategies described in this work will be of utmost importance for the forthcoming construction of solid-state-based quantum networks.

## METHODS

### Entangled photon source

The polarization-entangled photons are generated by a single GaAs QD embedded in a crystalline matrix of $Al_{0.4}Ga_{0.6}As$. The sample was fabricated on a GaAs (001) substrate using molecular beam epitaxy and the Al droplet etching technique[58], as in other recent implementations of quantum communication protocols[23,24]. The growth parameters were chosen to minimize the degree of in-plane anisotropy of the nanostructures[59]. The QDs were positioned at the middle of a 123 nm-thick layer of $Al_{0.4}Ga_{0.6}As$ placed between two 60 nm-thick layers of $Al_{0.2}Ga_{0.8}As$, which together constitute a λ-cavity. This cavity was inserted in between two distributed Bragg reflectors, composed by alternating layers of 70 nm of $Al_{0.95}Ga_{0.05}As$ and 60 nm of $Al_{0.2}Ga_{0.8}As$. The bottom part was formed by 9 pairs of layers, whereas the upper part was formed by 2 pairs of layers to facilitate light collection from the top sample surface. The sample was capped by 4 nm of GaAs to avoid surface oxidation. Due to the Bragg reflectors and a half-ball lens made of N-LASF9 placed on top of the sample, an extraction efficiency of approximately 12% is achieved for emission wavelengths around 785 nm.

### Quantum teleportation setup

The entangled photon source operates at a temperature of 5 K in a low-vibration closed-cycle He cryostat from attocube systems. The QD is driven via a resonant two-photon excitation scheme[9] employing a Ti:Sapphire femtosecond laser with an 80 MHz repetition rate. The rate of laser pulses is effectively doubled using an unbalanced Mach–Zehnder interferometer with a delay of 6.25 ns. Another similar interferometer creates consecutive laser pulses with 1.8 ns delay, as required for synchronous photon detection from a single source in the BSM and HOM setup. A 4f pulse shaper with an adjustable slit on its Fourier plane is used to reduce the linewidth of the laser down to approximately 200 µeV.

A 0.81 NA objective placed inside the cryostat focuses the laser on the sample and collects the signal from the QD. The backscattered light from the laser is removed by volume Bragg gratings notch filters with a bandwidth of 0.4 nm. A second set of volume Bragg gratings allows us to efficiently single out the emission from each of the two transitions of the XX-X cascade and reflect them into different parts of the setup. A liquid crystal retarder is used to compensate for any unitary transformation of the photon polarization induced by the optical elements, so that the two-photon polarization state retains maximal fidelity to the entangled state $|\phi^+\rangle$. Photons from the X to ground state transition are sent to the section of the setup devoted to the BSM. When specified, an air-spaced etalon with a bandwidth of 1.28 GHz and a free spectral range of 15 GHz is inserted before the BSM setup. The etalon is tilted in such a way that the signal transmission is maximal and equal for all its polarization components. An unbalanced Mach–Zehnder interferometer with a delay of 1.8 ns is inserted to match the delay introduced in the excitation path between consecutive laser pulses. Its second beam splitter is single-mode fiber-coupled with 48% reflectance, 52% transmittance and a mode overlap of 96%. Each of the outputs leads to a polarizing beam splitter. Two paddle (bat-ear) polarization controllers and a set of three zero-order waveplates, in the usual quarter-wave, half-wave and quarter-wave retardance combination, are used to preserve any input polarization state through the single-mode fibers. Photons from the XX to X transition go through a zero-order waveplate, with either quarter-wave or half-wave retardance, and a polarizing beam splitter to perform single-qubit quantum state tomography. A silicon avalanche photodiode with a time jitter of approximately 400 ps and a nominal quantum detection efficiency of 65% is placed at every output of each polarizing beam splitter—six in total. The single-photon counts are recorded by time tagging electronics with 10 ps resolution from Swabian Instruments and then analyzed to yield threefold coincidences histograms for different combinations of channels. A complete sketch of the experimental setup is reported in the Supplementary Information.

The autocorrelation function of the XX and X emission, the XX-X polarization density matrix and the HOM visibility of X photons from consecutive emission events are measured in the same setup. We estimate the HOM visibility from the ratio of the integrated intensity of the zero-time delay peak with respect to its side peaks in the measured coincidences histogram[60]. The peaks are modeled and fitted with five Gaussian functions convoluted with a symmetric exponential decay—which represent the instrument response function and the lifetime of the emitter respectively—with fixed width and different intensities. The result of the fitting procedure on the data is directly used to evaluate the integrated intensity of the side peaks. The zero-time delay peak has, in general, a more complex functional form due to the presence of quantum beats[61], so that the fit function does not fully reproduce its features even when the quantum beats are hidden by the limited time resolution. We calculated the integrated intensity of the central peak from the experimental data directly after having subtracted the fitted area of the side peaks.

Another silicon avalanche photodiode with time jitter slightly below 300 ps (FWHM) is used for lifetime measurements. The emission spectra of the QD are acquired by a deep-depletion, back-illuminated $LN_2$-CCD camera using a 750 mm focal length spectrometer, equipped with an 1800 g/mm grating.

### Data and error analysis

The errors in the estimation of the experimental fidelities are calculated by Gaussian propagation assuming a Poissonian distribution of the count of threefold coincidence events.

During the teleportation experiments, the data is recorded in two equivalent ways depending on the employed setup configuration. For the case without the spectral filtering of the etalon, we acquire a complete list of timestamps from all channels and use them to construct the threefold coincidences histograms within a correlation time of 100 ns. For the measurements performed with the etalon filter, due to the longer integration time and the large amount of collected timestamps, the detection events are analyzed in real-time and reduced to the threefold coincidences within a maximum time delay of 0.6 ps. The temporal window is chosen in a way that the fraction of discarded potential teleportation events is negligible, while the accidental threefold coincidences from side peaks are minimized. Eight combinations of channels are recorded, stemming from the two possible outputs of the quantum state tomography on the XX photons and from the four possible ways of probing the measurement of a Bell state on the X photons, two for the state $|\psi^-\rangle$ and the other two for $|\psi^+\rangle$. The polarization density matrices are built using the direct inversion single-qubit tomography method[49], which always returns a physical state for the data collected in our experiment.

## ACKNOWLEDGMENTS

This work was financially supported by the European Research Council (ERC) under the European Union's Horizon 2020 Research and Innovation Programme (SPQRel, grant agreement no. 679183), and the Linz Institute of Technology (LIT) through the LIT Secure and Correct Systems Lab funded by the state of Upper Austria and the Austrian Science Fund (FWF): P29603.

## AUTHORS' CONTRIBUTIONS

F.B.B. and F.S. performed measurements and data analysis with help from R.T.. L.S., K.D.J. helped with investigating the role of the etalon and with initial measurements. M.B.R., D.T. and E.R. contributed to the preparation of the setup. S.F.C.d.S., and A.R. designed and grew the sample. F.B.B., F.S and R.T. wrote the manuscript with feedback from all authors. All the authors participated in the discussion of the results. R.T. conceived the experiments and coordinated the project.

## COMPETING INTERESTS

The authors declare no competing financial interests.